\documentclass[%
reprint,
superscriptaddress,
amsmath,amssymb,
aps,
]{revtex4-2}

\usepackage{graphicx}
\usepackage{color}
\usepackage{dcolumn}
\usepackage{bm}
\newcommand{\tcb}[1]{\textcolor{blue}{#1}}

\begin{document}

	\title{Discovery of the Solution to the ``Einstein-Podolsky-Rosen Paradox''} 
	\author{Roman Schnabel}
	\email{roman.schnabel@uni-hamburg.de}
	\affiliation{%
		Institut f\"ur Quantenphysik und Zentrum f\"ur Optische Quantentechnologien\\
		Universit\"at Hamburg, Luruper Chaussee 149, 22761 Hamburg, Germany}

\date{\today}

\begin{abstract} 
In 1935, Albert Einstein, Boris Podolsky and Nathan Rosen (EPR) published a thought experiment that is entirely correct, has been demonstrated in real experiments, and is now the most famous in quantum physics. Their pioneering work described, for the first time, quantum correlations and can be regarded as a very early glimpse into today's `deep' quantum technologies, by which I mean those that enhance functionality by making use of quantum correlations. However, their work also produces a paradox that Erwin Schr\"odinger had already recognised as such in 1935 and which has since been cemented by the so-called Bell experiments. Here, I can now pinpoint the cause of the paradox in the EPR line of reasoning. 
\tcb{It boils down to a single assumption by EPR, which I prove to be false.} I show that this resolves the paradox.
\end{abstract}

\maketitle

\section*{Introduction}\vspace{-2mm}
Quantum theory (QT) is arguably the most successful theory given the vast number of different kinds of experiments that are correctly described. Not a single experiment contradicts QT. Given this, QT is \tcb{widely} regarded as a correct and complete theory (in flat space time).
When the QT was formulated in the second half of the 1920s, its later success was not clear. One of its most prominent detractors was Albert Einstein. Together with his two co-authors Boris Podolsky and Nathan Rosen, he brought his criticism to the point in 1935, when they published a manuscript entitled \textquotedblleft \emph{Can Quantum-Mechanical Description of Physical Reality Be Considered Complete?}\textquotedblright \cite{Einstein1935}, which is today one of Einstein's most cited works. 
This work is probably the first to examine measurement results obtained from two entangled systems. (The term `entanglement' was coined by Erwin Schr\"odinger \cite{Schroedinger1935}.) It has inspired a vast number of scientific papers, including the formulation of Bell's inequalities and their experimental testing. A modern description is that the EPR paper examines the correlations between measurement results obtained from two uncertainty regions, which the correlations are more precise than the minimum uncertainty region according to Heisenberg's uncertainty principle \tcb{\cite{Reid1989}}. By exploiting quantum correlations, today's gravitational wave detectors can peer deeper into the Universe than ever before \cite{LSC2011,Grote2013,Tse2019,Acernese2019,GWo3a2021} and in the future, they are expected to enable quantum computers to achieve the necessary fault tolerance \cite{Shor1996}.

To investigate the question of the completeness of QT, \tcb{EPR draw almost entirely on QT itself}. In particular,
they propose a thought experiment, which is about measurements on a quantum-entangled system. Their aim was to determine whether measurements taken within regions of uncertainty yield values that already existed prior to the measurement, or whether the measurement results arise truly randomly  without any form of causality.

EPR tried to find out the answer to their question by formulating a \textquotedblleft {reasonable}\textquotedblright~sufficient criterion for  \textquotedblleft physical reality\textquotedblright:
\begin{quote}\vspace{-1mm}
\textquotedblleft \emph{If, without in any way disturbing a system, we can predict with certainty (i.e.,\;with probability equal to unity) the value of a physical quantity, then there exists an element of reality corresponding to that quantity.}\textquotedblright \hspace*{\fill} 
\end{quote}\vspace{-1mm}
Fortunately, we do not need to discuss their definition of `reality', because the same column of the first page of \,\cite{Einstein1935} provides another sufficient criterion. If we combine both criteria, the term `reality' is bypassed and we get the actual criterion that EPR considered sufficient for proving incompleteness of a physical theory. This leads to a statement which I would like to refer to as the `EPR implication'.
\begin{quote}\vspace{-1mm}
\textquotedblleft \emph{If, without in any way disturbing a system, we can predict with certainty (i.e.,\;with probability equal to unity) the value of a physical quantity --- but this value has no counterpart in the physical theory --- then the theory is incomplete.}\textquotedblright \hspace*{\fill} $[\!\;{\rm I}\!\;]$
\end{quote}\vspace{-1mm}
\tcb{EPR did not prove this implication. Nor is it a postulate of QT. This is an {\it assumption} made by EPR as part of their analysis. For EPR, its correctness follows from pure logic. In their search for the answer to the question posed in the title of \cite{Einstein1935}, implication $[\!\;{\rm I}\!\;]$ is the only element that does not stem from QT.}
EPR continued their publication with the presentation of their thought experiment. It was a special, physically correct scenario, in full accordance to QT, and it surprisingly fulfilled all conditions of the above implication $[{\rm \!\;I\!\;}]$. 
Their publication concludes: 
\textquotedblleft \emph{While we have thus shown that $[{\rm QT}]$ does not provide a complete description of the physical reality, we left open the question of whether or not such a description exists. We believe, however, that such a theory is possible.}\textquotedblright ~

Erwin Schr\"odinger and Niels Bohr promptly published their \tcb{comments} \cite{Bohr1935,Schroedinger1935}, 
\tcb{in which} they regarded QT as complete and did not agree with the final result of the EPR analysis, but were also unable to pinpoint where the error lay in the EPR's line of reasoning. 
\tcb{For Schr\"odinger, the EPR analysis 
revealed a ``{\it discomforting}'' property of QT and brought a paradox to light \cite{Schroedinger1935}. He described the paradox by deriving a seemingly paradoxical ``steering effect'' from the EPR thought experiment. 
I would like to describe Schr\"odinger's EPR paradox here as follows. 
According to QT, it is not possible to predict a precise measurement value within the uncertainty range of an observable of a single, isolated system. This means that the measurement value is truly random. However, the same QT allows the prediction of a precise measurement value within the uncertainty range of an observable of a system if the value of a precise corresponding measurement on a second system, which is entangled with the first, is known. 
This situation behaves as if the precise measured value from the first system were not truly random.}

EPR's work has attracted a lot of attention only after 1964 when John Bell \cite{Bell1964} succeeded in formulating an inequality for an experimental test, whether QT can be expanded and thus `completed'. His inequality was based on pure mathematics and did not assume the correctness of QT.
The experimental violation of Bell's inequality would prove that the seemingly missing counterparts (see implication $[\!\;{\rm I}\!\;]$) of the precisely predictable measurement values in the EPR thought experiment (usually called `local hidden variables') cannot exist, in general \cite{Clauser1969}.
\\
QT itself suggests a specific class of statistical experiments that should violate Bell's inequality. The experiments are often called `Bell tests'. Since the 1970s, up to recent years, these experiments have violated Bell's inequality \cite{Freedman1972,Aspect1981,Maddox1991,Kwiat1995,Weihs1998,Ansmann2009,Gisin2009,Giustina2013,Hensen2015,Shalm2015,Moreau2019}. 
The Bell experiments have thus (i) substantiated the possibility of predicting with certainty the exact measured value of a physical quantity, even though quantum mechanics (QM) has no counterpart to this value, by utilising the measured value from an entangled subsystem, and (ii) ruled out the possibility of finding a physical theory that is more complete than quantum mechanics in the manner envisaged by EPR.
The results of Bell tests shaped the statement \textquotedblleft {Nature lacks local realism\,}\textquotedblright~and is \textquotedblleft {non-local\,}\textquotedblright
\cite{Aspect1981,Ghosh1986,Weihs1998,Groblacher2007}. 
The results of Bell tests cemented the combination of (i) and (ii), which corresponds to the fundamental paradox already seen by Erwin Schr\"odinger in 1935. It is not resolved so far \cite{Maddox1991,Gisin2009}.

The paradox raised the question of whether the Bell tests might have loopholes. But to promote potential loopholes, strange and implausible assumptions about Nature have to be made. Anyway, all three main potential loopholes have since been closed by experiments \cite{Weihs1998,Giustina2013,Hensen2015,Shalm2015}. 
Conclusions (i) and (ii) drawn from the Bell tests have been thus again confirmed.
Since then, metaphysical models that manage to circumvent the paradox have gained strength. They can neither be proven nor disproven. Particularly noteworthy is Everett's many-worlds interpretation from 1957 \cite{Everett1957}, which is very popular in parts of the quantum physics community today.
\tcb{Both before and after Bell's work, it was proposed that QT could be completed by means other than local hidden variables. Bohmian mechanics \cite{Bohm1952} proposes {\it non-local} hidden variables, which would, however, lead to complete determinism in all processes in the Universe. Other studies assume the existence of truly random changes in the physical properties of systems, but without these systems interacting with the environment (`spontaneous localisation'). Bell himself put forward a proposal along these lines with his stochastic ``beables'' \cite{Bell2004}. 
However, the EPR thought experiment does fundamentally involve an interaction with the environment, in this case in the form of precise measurements carried out by an observer. Clearly, though, no actual person is required. The generalised measurement is the exchange of energy with the thermal bath. The higher the bath temperature, the more precise the measurement. 
In the absence of interaction with the environment, the Schr\"odinger equation \cite{Schroedinger1926a} predicts a deterministic evolution of the uncertainty ranges, specifically a continuously expanding position uncertainty.
In summary, it can be said that none of these proposals is satisfactory, as they raise new, unresolved questions.}

\tcb{Here I propose the solution to the EPR paradox, as it arose for Schr\"odinger and which persists to this day. To this end, I analyse in detail the EPR chain of reasoning, which they construct on the basis of QT in conjunction with their implication. I discover that the EPR chain of reasoning allows for an equivalent but linguistically refined EPR-implication. 
Utilising the recently published physical concept by which position-momentum-entangled free particles can be produced \cite{Schnabel2025epr}, I prove that the EPR implication $[\!\;{\rm I}\!\;]$ is generally incorrect, which localizes the specific error in the EPR paper. For readers who do not wish to follow this concept in detail, I present alpha decay as a meaningful analogue.
\\
Surprisingly, my work demonstrates the following: even if a measured value can be predicted with certainty, a theory does not necessarily need to include a counterpart for it in order to be complete. With my work, the combination of the possibility of accurately predicting a measured value and the absence of a corresponding counterpart in the theory becomes even logically consistent. 
This paradigm shift resolves Schr\"odinger's EPR paradox. 
For the following statement is clear from my work: {\it EPR's analysis of their thought experiment does not contradict the generally assumed property of QT that a precise measurement value within the uncertainty range of an observable is always truly random.}
Consequently, the EPR paper no longer provides any justification for the view that QT must be completed by local hidden variables. My work requires no additional mechanisms or assumptions regarding QT.}

\vspace{-2mm}
\section*{About the EPR thought experiment}
\vspace{-2mm}
If QT is correct, then it follows that the EPR thought experiment is correct. This statement is undisputed.
According to EPR's argument, their thought experiment shows, however, that QT describes the actually precise values of physical quantities incompletely. The presumed incomplete description relates to the quantum uncertainties.
Through their thought experiment, EPR considered it proven that quantum uncertainties are not fundamental, but that more precise values of physical quantities do in fact exist, only that they are not captured by QT.\\ 

The EPR thought experiment proceeds as follows. There is a system consisting of two freely moving particles, A and B. 
Before the measurement, the two particles are in an entangled state of motion according to QT. 
The experiment opens up two freely selectable options for the experimenter. Either they measure the two positions $\hat x_{\rm A}$ and $\hat x_{\rm B}$ (with respect to a common coordinate system) producing the data $(x_{{\rm A,}i=1}; x_{{\rm B,}i=1})$ or they measure the two momenta  $\hat p_{\rm A}$ and $\hat p_{\rm B}$ (with respect to the same coordinate system) producing the data $(p_{{\rm A,}j=1}; p_{{\rm B,}j=1})$.
Momentum may be approximated as the product of particle mass and velocity, because the incorrectness of the EPR implication will already become apparent in this limiting case of small velocities. The ``hat'' above $x$ and $p$ marks the expressions as physical quantities in order to distinguish them from the statistically distributed measurement values of the same.
Side note: The {distance} between the two particles and the {simultaneity} of the two measurements would only need to be discussed with regard to potential loopholes of Bell tests, but not with regard to the objective of this work.\\
The measurements must now be repeated many times in relation to the same coordinate system with fresh, identically entangled systems so that standard deviations of the measured values can be calculated. 
The word `identical' is important here. For the EPR thought experiment to work, all measured systems (consisting of two particles) and the entangled state of motion in which the system finds itself must be indistinguishable within the formalism of QT. The possibility of indistinguishability is a direct consequence of quantization. In experiments, achieving  indistinguishability according to QT is feasible. 

In roughly half of the measurements, the experimenter should have measured the positions and in the other half the momenta. The standard deviation $\Delta \hat x_{\rm A}$ of the measured values $\{x_{{\rm A,}i} \}$ is defined as the square root of the variance $\Delta\!^2 \hat x_{\rm A} = \langle \hat x_{\rm A}^2 \rangle - \langle \hat x_{\rm A} \rangle^2$, where the brackets denote the mean value, also called the `expectation value' in QT.

For all measurements on particle A (or particle B), Heisenberg's uncertainty principle applies as part of QT individually for each particle
\cite{Heisenberg1927,Kennard1927,Weyl1927,Robertson1929}
\begin{equation}
\Delta \hat x \cdot \Delta \hat p \geq \hbar/2 \, ,
\label{eq:1}
\end{equation}
where $\hbar \approx 1.0546 \times 10^{-34}$Js is the reduced Planck constant, a constant of Nature \cite{Planck1900}. 
Analysis of the measurement data from the ERR thought experiment shows that the inequality is more than satisfied. (Side note: The stronger the entanglement, the further the uncertainty product departs from the minimum.)
The interesting point about the above inequality is that the uncertainty product cannot be smaller than a certain value ($\hbar/2$) that is greater than zero. Eq.\,(\ref{eq:1}) leads to three possible conclusions, which have been repeatedly discussed in the decades since the QT was formulated.\\[2mm]
(a) A measurement statistic obtained may be based on absolutely identical measurements on absolutely identical systems, but different measurement values still arise on the basis of true randomness, i.e., without any causality.
In this case, no theory can causally link the measured values, contain counterparts, or make precise predictions.
\\[2mm]
(b) Although a measurement statistic obtained may be based on absolutely identical measurements on seemingly absolutely identical systems according to QT, the different measured values illustrate that the systems are in fact distinguishable.\\[2mm]
(c) Although a measurement statistic obtained may be based on seemingly absolutely identical measurements on absolutely identical systems, the different measured values illustrate the fundamental influence of the measurement apparatus.\\[2mm]
With the formulation of QT in the second half of the 1920s \cite{Heisenberg1925,Born1925,Born1926,Schroedinger1926, Heisenberg1927,Bohr1928e}, conclusion (a) was rightly the prevailing view among the founders of quantum physics. The work of Einstein, Podolsky, and Rosen \cite{Einstein1935} explicitly opposed this prevailing view and believed it could justify why (b) should be the correct view. View (c) rightly played only a minor role in the discussions. 
From today's perspective, (c) has indeed been refuted by the successful realizations of the EPR thought experiment.\\

Let's move on to the successful execution of the EPR thought experiment.
As described above, it provides a large number of measured data pairs  
$\{(x_{{\rm A,}i}; x_{{\rm B,}i})\}$ and $\{(p_{{\rm A,}j}; p_{{\rm B,}j})\}$.
They allow a statistical analysis of the individual sums and differences $(x_{{\rm A,}i} \pm x_{{\rm B,}i})$, $(p_{{\rm A,}j} \pm p_{{\rm B,}j})$, respectively.
According to EPR (and according to QT) the statistical analysis approaches one of the two following extreme cases
\begin{align} \label{eq:2}
	\Delta (\hat x_{{\rm A}} - \hat x_{{\rm B}})  =  \Delta (\hat p_{{\rm A}} + \hat p_{{\rm B}}) &= 0 \hspace{5mm} \\
{\rm or} \hspace{5mm}	
	\Delta (\hat x_{{\rm A}} + \hat x_{{\rm B}})  =  \Delta (\hat p_{{\rm A}} - \hat p_{{\rm B}}) &= 0 \, , 
\label{eq:3}
\end{align} 
depending of the sign of the quantum correlation in the position-momentum-entangled state of particles A and B. 
If the initial statistical analysis of the EPR thought experiment has confirmed the validity of either equation (2) or (3), it is possible, upon continuing the series of measurements, to predict {\it predict every individual position value at B and every individual momentum value at B with absolute precision if the same quantity has been measured at A}.\\
If the implication $[{\rm \!\;I\!\;}]$ were correct, one would have to conclude, like EPR, that the above conclusion (b) is correct and that  systems that are identical within the framework of QT are in fact different. The difference would thus imply the existence of `hidden variables', and QT would be incomplete because it does not know about these variables.
It should be noted that Eq.\,(\ref{eq:1}), (\ref{eq:2}), and Eq.\,(\ref{eq:3}) can all be derived within the framework of QT, namely from one and the same equation, and in a very general way, independent of the states at which measurements are taken.\\

Real experiments have long since fully confirmed the correctness of the EPR thought experiment. (The measured quantities were formally equivalent but not identical to the positions and momenta of two freely moving particles.) The first studies demonstrated a modified version of the EPR thought experiment and aimed to test Bell's inequality. They used measurement quantities with a discrete value spectrum, namely photon numbers in linearly polarized modes of fluorescence light \cite{Clauser1969,Freedman1972}. 
The first experiment that performed the EPR thought experiment in such a way that it was described by a Heisenberg uncertainty relation analogous to (\ref{eq:1}) also confirmed the EPR thought experiment as correct \cite{Ou1992}. 
Measured were the entangled dimensionless amplitude modulation depths $\hat X_{{\rm A,B}}$ and phase modulation depths $\hat Y_{{\rm A,B}}$ of the electric field strengths of the light waves of two laser beams, which are described by {\cite{Reid1989,Schnabel2017}}
\begin{align} 
\label{eq:4} 
	\Delta \hat X \cdot \Delta \hat Y &\geq 1 \, ,
\end{align} 
where the normalization is such that the ground state with zero photons has the uncertainties
\begin{align} 
\label{eq:5} 
	\Delta \hat X_0 = \Delta \hat Y_0 = 1 \, .
\end{align} 
From each measurement value of $\hat X_{{\rm A}}$ (or $\hat Y_{{\rm A}}$) in \cite{Ou1992}, the corresponding measurement value of $\hat X_{{\rm B}}$ (or $\hat Y_{{\rm B}}$) was inferred with an imprecision that was smaller than B's ground state uncertainty, i.e., $\Delta \hat X_{\rm B,inf}, \Delta \hat Y_{\rm B,inf} \!<\! 1$. 
The EPR paradox was demonstrated with the quantitative measure \tcb{\cite{Reid1989}}  $\Delta\!^2 \hat X_{\rm B,inf} \cdot \Delta\!^2 \hat Y_{\rm B,inf} \approx 0.7 < 1$. 
EPR entanglement of this type was subsequently used for unconditional quantum teleportation \cite{Furusawa1998, Bowen2003, Bowen2003b}. The strongest EPR entanglement demon\-strated to date, analogous to \cite{Ou1992}, is $\Delta\!^2 \hat X_{\rm B,inf} \cdot \Delta\!^2 \hat Y_{\rm B,inf} \approx 0.0309 < 1$  \cite{Eberle2013}.
 Entanglement of quantities described by a Heisenberg uncertainty relation has also been successfully demonstrated with atomic many-body systems. In \cite{Julsgaard2001}, the entanglement of two atomic clouds of approximately $10^{12}$ atoms at room temperature was demonstrated with respect to two non-commuting projections of their collective spins. In \cite{Peise2015}, a small fraction of a Bose-Einstein condensate of approximately $2 \times 10^{4}$ atoms was brought pairwise into two symmetrically split hyperfine levels. Observables according to inequality (\ref{eq:4}) were defined, and the EPR thought experiment was realized with these.

\vspace{-2mm}
\section*{EPR's line of reasoning} 
\vspace{-2mm}
{The Implication $[\!\;{\rm I}\!\;]$ is given \tcb{on the first page of the EPR paper \cite{Einstein1935}.
On the subsequent pages, EPR outline the properties of QT and place them in context.}
}
The text on the second page essentially says the following. 
{\it If the momentum of a free particle has a certain value $p_0$ [...], then all values of the particle position are equally probable.} 
Here, the true randomness with which measured values arise within the boundary set by Heisenberg's uncertainty principle is described.
Furthermore, one finds: {\it ``The usual conclusion from this in quantum mechanics is that when the momentum of a particle is known, its coordinate [position] has no physical reality''}.
Here it becomes clear that EPR understands ``a quantity without physical reality'' to be a quantity whose measured values are truly random, i.e., occur without causal reason.   
Accordingly, EPR's reasoning for their implication can be added to $[\!\;{\rm I}\!\;]$  as follows.
\begin{quote}\vspace{-2mm}
\emph{...since a value that is predictable cannot be the result of a truly random process, which would in fact be the only process whose emerging value is outside the realm of causality and physical theory.}
\end{quote}\vspace{-1mm}
The last half of this statement is in line with the thinking of contemporary quantum physicists and is undoubtedly correct. It forms the basis for the fact that (a) is a possible conclusion that can be drawn from Heisenberg's uncertainty principle, see text beneath inequality (\ref{eq:1}).

\tcb{Based on this analysis, the EPR implication can be rephrased as follows without altering its meaning
\begin{quote}\vspace{-1mm}
\textquotedblleft \emph{If, without in any way disturbing a system, we can predict with certainty (i.e.,\;with probability equal to unity) the value of a physical quantity, then the measured value is not the result of a truly random process, i.e., it does not lie outside the realm of causality.}\,\textquotedblright \hspace*{\fill} $[\!\;{\rm II}\!\;]$
\end{quote}\vspace{-1mm}
}
The final question I am considering is whether this sentence is correct.

\vspace{-2mm}
\section*{The solution to the EPR paradox}
\vspace{-2mm}

To answer the final question, \tcb{I consider the {\it generation} of the position-momentum-entanglement that is required for the original EPR thought experiment.} Remarkably, EPR \cite{Einstein1935} did not describe the mechanism of the entanglement generation. It was only recently discovered \cite{Schnabel2025epr}.
Accordingly, a simple, \tcb{idealised} one-dimensional elastic collision between two \tcb{force-free} particles is sufficient to generate the original EPR entanglement, provided that they have an unbalanced mass ratio of, for example, 1:3, as well as separable initial states that possess a squeezed position uncertainty and a squeezed momentum uncertainty, respectively.
EPR's position-momentum-entanglement arises then from the redistribution of the initially separable quantum uncertainties under the conditions of energy and momentum conservation. 
\\
\tcb{
The mechanism for generating the original EPR entanglement \cite{Schnabel2025epr} makes it evident that all properties of the measurement values in the EPR thought experiment are consistent within the framework of QT, and that no further assumptions need to be made.
Firstly, it is evident that the entanglement following the elastic collision persists for an extended period of time, without this giving rise to any conflict with the complete absence of any forces between A and B.
Secondly, it is evident that after the elastic collision neither A nor B has a precisely defined position or momentum prior to a measurement (if Heisenberg’s uncertainty principle is correct). It follows directly from this that any measured value must be truly random in its deviation from the expectation value (of classical mechanics).
Thirdly, it is evident that, due to the states being squeezed at the outset, a measured value at A always allows for a precise inference (`prediction') of the corresponding measured value at B, i.e., Eq.\,(\ref{eq:2}) is fulfilled.
}
For the limit of infinitely high squeeze values prior to the elastic collision, the measurements correspond to Eqs.\,(\ref{eq:2}) or (\ref{eq:3}).
\tcb{
The combination of these three facts conclusively demonstrates that the EPR implication (in the forms $[\!\;{\rm II}\!\;]$ and $[\!\;{\rm I}\!\;]$) is incorrect. (The word `predict' in the EPR implication calls for a comment. EPR would have been better off choosing the term `infer' instead, as used in \cite{Reid1989}, because when one subsystem is decoupled from the entangled state through decoherence (the measurement), this also happens instantaneously to the other. The `prediction' therefore takes place {\it after} a definite position or momentum has been randomly realised for both subsystems.)
}
\\
\tcb{
My analysis begins at the same point as that of EPR, with exactly the same thought experiment and assumption that quantum theory is correct and complete. This includes the validity of Heisenberg’s uncertainty principle, including the consequence that precise measurements of regions of uncertainty yield truly random results.
Unlike EPR, I find that QT is internally consistent. No EPR-style contradiction arises, and therefore there is neither reason to suspect that QT is incomplete \cite{Einstein1935} nor, alternatively, to see a paradox \cite{Schroedinger1935}. The result of the analyses differ because EPR assumes the correctness of their implication, whilst I use the generation of EPR entanglement \cite{Schnabel2025epr} to prove that}
the EPR implication is false.
\\

\tcb{
The above analysis of the EPR thought experiment requires an understanding of how Heisenberg’s uncertainty principle inevitably leads to the occurrence of random measurement results. 
To get round this, I shall discuss here another physical experiment which is considerably simpler, but which can be used to directly refute the EPR implication. This is the observation of alpha decay, as in this case it is easy to accept the existence of true randomness.
\\
}
We consider the radioactive alpha decay of a large number of the same suitable nucleus, e.g., uranium-238.
A characteristic quantity of the type of nucleus is the half-life. If you start observing a single nucleus at any point in time, there is a genuine, equally weighted quantum mechanical randomness as to whether the atom decays before the half-life expires or only afterwards. If you interrupt the observation for any length of time and then find that the atom has not yet decayed, you are back in exactly the same situation as at the start of the first observation phase.
The uranium nucleus does not undergo any ageing process throughout its existence. Consequently, there is no internal temporal evolution leading to decay. The precise decay time of the individual nucleus is thus not captured by the QT. Such a counterpart is missing but the QT complete.

\tcb{
A uranium-238 nucleus exists in an entangled state that encompasses two simultaneously possible outcomes: either a thorium-234 nucleus and an alpha particle have taken the place of the uranium-238 nucleus, or the latter still exists. 
The entanglement presumably arose a long time ago during the merger of two neutron stars. Since then, there has been no interaction between a thorium-234 nucleus and an alpha particle within the uranium nucleus, as neither particle exists. At some point, alpha decay occurs as a truly random event. 
}
It happens without cause and without any trigger, which is referred to as `spontaneous' in quantum physics. 
Two decay products are formed simultaneously: a helium nucleus (alpha particle) and a nucleus of an element that is lighter than the element of the original nucleus, here, thorium-234. 
When an ideal alpha-particle detector with perfect efficiency witnesses the decay by {\it instantaneously} detecting the alpha particle, it is 100\% certain that a thorium-234 appeared as well in that very moment. Both new nuclei emerged jointly in a truly random way.
(I would point out that the situation is different from that of Bertlmann's Socks \cite{Bell2004}, which are observed by a human long after the thermal bath of the environment has already been exchanging energy with the socks, i.e.,\,``{\it disentangled\,}'' the socks \cite{Schroedinger1935}.)
The alpha decay can actually be understood as two truly random processes that are precisely correlated with each other due to a boundary condition, first of all conservation of energy (including mass). That is what we today refer to as ``quantum correlation''.
My conclusion is therefore:
\begin{quote}\vspace{-2mm}
\emph{The predictability of a measured value does not rule out the possibility that the value is the result of a truly random process!}
\end{quote}\vspace{-1mm}
The corrected analysis of the EPR thought experiment and, independently, the alpha decay do nothing more and nothing less than prove that the core of EPR's line of reasoning is incorrect. They localize the error in the EPR implication, which Einstein {\it et al.} erroneously derived from their thought experiment.
With the alpha decay in mind, it becomes logical that a truly random value can still be predicted precisely if a second, equally random value appears in parallel that correlates perfectly with the first value. This is precisely the situation in the EPR thought experiment and in all Bell tests.\\

One could criticise the example of alpha decay on the grounds that it only has {\it one} measurement variable---namely the existence of a particle---whereas the EPR thought experiment uses two measurement variables, namely position and momentum of a particle. 
\tcb{However, this criticism would not apply in the context of this manuscript.
Let us suppose, for a change, that one were to modify the EPR thought experiment and not carry out any momentum measurements, but instead set in advance that, prior to a (position) measurement, neither particle has a defined position. In this scenario, EPR would again correctly argue that this situation -- in principle -- is possible according to QT, as evidenced by the fact that QT does not include a counterpart to position for this scenario prior to a measurement. They would go on to argue, incorrectly, that the thought experiment of the same QT refutes this, because following the position measurement on A, the result of the position measurement on B is always precisely predictable, even if the measurement on A does not cause any disturbance to B. EPR would therefore conclude---even with a {\it single} observable---that B already has a defined position prior to the measurement at A, for which the counterpart is missing in the QT.
}
The EPR thought experiment actually \tcb{uses} position measurements {\it and} momentum measurements in order to \tcb{employ} the limit beyond which the precision of the measured values means that QT no longer has any counterparts for them, i.e., when true randomness has to be present according to QT. This is the case when the predictions of the positions and momenta of system B violate inequality (\ref{eq:1}). 
The EPR thought experiment is thus more rigorously grounded in quantum theory, but two commuting observables are not necessary for the conclusion that EPR derive from it.
\tcb{The example of alpha decay can serve its didactic purpose by first (and quite rightly) establishing that the decay of a single atomic nucleus, whether before or after the half-life, is genuinely random and therefore has no counterpart in QT. It is clear here that the detection of an alpha particle precisely predicts the existence of a thorium nucleus.
}

\vspace{-0mm}
\section*{Summary and conclusions}
\vspace{-2mm}
\tcb{The EPR paper \cite{Einstein1935} is a foundational publication in modern physics underpinning the field of the quantum
information. It is also true, however, that the EPR paper reaches the following incorrect conclusion:
``{\it ...we have thus shown that the wave function [quantum theory] does not provide a complete description of the physical reality...}''.
EPR clearly describe the nature of the incompleteness they perceive in QT. This is precisely the type of incompleteness addressed by the Bell experiments. Assuming that the Bell experiments contain no loophole, the EPR conclusion is proven to be incorrect. EPR must therefore have made a mistake in their analysis, or quantum physics presents us with a paradox \cite{Schroedinger1935}.}
\\
\tcb{In my work, I resolve the EPR paradox by pinpointing the flaw in the EPR paper. I identify the sole assumption made by EPR and refute it. This is the following assumption, which I call the EPR implication and which I repeat here in full but in a linguistically abbreviated form: ``{\it Precisely predictable measurement results cannot be the result of a truly random process.}''
I refute this statement by referring to the process for generating position-momentum-entangled free particles \cite{Schnabel2025epr}. Furthermore, I cite the well-known alpha decay of the uranium-238 nucleus as an analogous example, in which it is evident that the realisation of an alpha particle by a perfect measuring apparatus precisely predicts the realisation of a thorium-234 nucleus. The quantum-mechanical realisations of the alpha particle and the thorium nucleus are genuinely random, but the perfect correlation between the realisations of the particles makes the possibility of prediction obvious.}
\\
Historically, the work of EPR \cite{Einstein1935} was the motivation for deriving Bell inequalities and for conducting Bell tests. The error in the EPR paper means that this motivation did not exist in the first place. 
However, this does not diminish the today's deep fundamental significance of the tests of Bell's inequality that have been carried out. 
\tcb{It is probably reasonable to assume that QT does not involve any other form of incompleteness either. 
At any rate, there are no convincing arguments to the contrary. In this case,} experimental violations of Bell's inequalities have proven that the exchange of energy between physical systems can lead to genuinely random values for the properties of those systems.
\tcb{The highly remarkable possibility that a truly random result, one without any cause or trigger, can be proven as such remains untouched despite the discovery of the error in the EPR paper.}\\

\vspace{0mm}
\begin{acknowledgments}\vspace{-2mm}
This work was performed within the European Research Council (ERC) Project \emph{MassQ} (Grant No.~339897) and within Germany's Excellence Strategy -- EXC 2056 `Advanced Imaging of Matter', project ID 390715994 and EXC 2121 `Quantum Universe', project ID 390833306, which are financed by the Deutsche Forschungsgemeinschaft.
RS thanks Serge Reynaud for many fruitful discussions.
\end{acknowledgments}
\vspace{4mm}

\section*{Disclosures}\vspace{-3mm}
The authors declare no conflicts of interest.\\[-2mm]


\begin{thebibliography}{10}

\bibitem{Einstein1935}
A.~Einstein, B.~Podolsky, and N.~Rosen.
\newblock {Can Quantum-Mechanical Description of Physical Reality Be Considered
  Complete?}
\newblock {\em Physical Review}, 47(10):777--780, may 1935.

\bibitem{Schroedinger1935}
E.~Schr{\"{o}}dinger.
\newblock {Discussion of Probability Relations between Separated Systems}.
\newblock {\em Mathematical Proceedings of the Cambridge Philosophical
  Society}, 31(4):555--563, oct 1935.

\bibitem{Reid1989}
M~Reid.
\newblock {Demonstration of the Einstein-Podolsky-Rosen paradox using
  nondegenerate parametric amplification}.
\newblock {\em Physical Review A}, 40(2):913--923, jul 1989.

\bibitem{LSC2011}
J.~Abadie~\emph{et al.}
\newblock {A gravitational wave observatory operating beyond the quantum shot-noise limit}.
\newblock {\em Nature Physics}, 7(12):962--965, sep 2011.

\bibitem{Grote2013}
H.~Grote, K.~Danzmann, K.~L. Dooley, R.~Schnabel, J.~Slutsky, and H.~Vahlbruch.
\newblock {First Long-Term Application of Squeezed States of Light in a Gravitational-Wave Observatory}.
\newblock {\em Physical Review Letters}, 110(18):181101, may 2013.

\bibitem{Tse2019}
M.~Tse~\emph{et al.}
\newblock {Quantum-Enhanced Advanced LIGO Detectors in the Era of
  Gravitational-Wave Astronomy}.
\newblock {\em Physical Review Letters}, 123(23):231107, dec 2019.

\bibitem{Acernese2019}
F.~Acernese~\emph{et al.}
\newblock {Increasing the Astrophysical Reach of the Advanced Virgo Detector
  via the Application of Squeezed Vacuum States of Light}.
\newblock {\em Physical Review Letters}, 123(23):231108, dec 2019.

\bibitem{GWo3a2021}
R.~Abbott~et al.
\newblock {GWTC-2: Compact Binary Coalescences Observed by LIGO and Virgo
  during the First Half of the Third Observing Run}.
\newblock {\em Physical Review X}, 11(2):021053, 6 2021.

\bibitem{Shor1996}
P. W. Shor.
\newblock {Fault-tolerant quantum computation}.
\newblock In {\em Proceedings of 37th Conference on Foundations of Computer
  Science}, pages 56--65. IEEE Comput. Soc. Press, 1996.

\bibitem{Bohr1935}
Niels Bohr.
\newblock {Can Quantum-Mechanical Description of Physical Reality be Considered Complete?}
\newblock {\em Physical Review}, 48(8):696--702, oct 1935.

\bibitem{Bell1964}
John~S. Bell.
\newblock {On the Einstein Podolsky Rosen Paradox}.
\newblock {\em Physics}, 1:195--200, 1964.

\bibitem{Clauser1969}
John~F. Clauser, Michael~A. Horne, Abner Shimony, and Richard~A. Holt.
\newblock {Proposed experiment to test local hidden-variable theories}.
\newblock {\em Physical Review Letters}, 23:880, 1969.

\bibitem{Freedman1972}
Stuart~J. Freedman and John~F. Clauser.
\newblock {Experimental Test of Local Hidden-Variable Theories}.
\newblock {\em Physical Review Letters}, 28(14):938--941, apr 1972.

\bibitem{Aspect1981}
Alain Aspect, Philippe Grangier, and Gerard Roger.
\newblock {Experimental Tests of Realistic Local Theories via Bell`s Theorem}.
\newblock {\em Physical Review Letters}, 47:460, 1981.

\bibitem{Maddox1991}
John Maddox.
\newblock {Non-locality bursts into life}.
\newblock {\em Nature}, 352:277--279, 1991.

\bibitem{Kwiat1995}
Paul~G Kwiat, Klaus Mattle, Harald Weinfurter, Anton Zeilinger, and Alexander~V
  Sergienko.
\newblock {New High-Intensity Source of Polarization-Entangled Photon Pairs}.
\newblock {\em Physical Review Letters}, 75:4337, 1995.

\bibitem{Weihs1998}
Gregor Weihs, Thomas Jennewein, Christoph Simon, Harald Weinfurter, and Anton
  Zeilinger.
\newblock {Violation of Bell's Inequality under Strict Einstein Locality
  Conditions}.
\newblock {\em Physical Review Letters}, 81(23):5039--5043, dec 1998.

\bibitem{Ansmann2009}
Markus Ansmann, H.~Wang, Radoslaw~C. Bialczak, Max Hofheinz, Erik Lucero,
  M.~Neeley, A.~D. O'Connell, D.~Sank, M.~Weides, J.~Wenner, A.~N. Cleland, and
  John~M. Martinis.
\newblock {Violation of Bell's inequality in Josephson phase qubits}.
\newblock {\em Nature}, 461(7263):504--506, 2009.

\bibitem{Gisin2009}
Nicolas Gisin.
\newblock {Quantum nonlocality: How Does Nature Do It?}
\newblock {\em Science}, 326:1357--1359, 2009.

\bibitem{Giustina2013}
Marissa Giustina, Alexandra Mech, Sven Ramelow, Bernhard Wittmann, Johannes
  Kofler, J{\"{o}}rn Beyer, Adriana Lita, Brice Calkins, Thomas Gerrits,
  Sae~Woo Nam, Rupert Ursin, and Anton Zeilinger.
\newblock {Bell violation using entangled photons without the fair-sampling
  assumption.}
\newblock {\em Nature}, 497(7448):227--30, may 2013.

\bibitem{Hensen2015}
B.~Hensen, H.~Bernien, A.~E. Dr{\'{e}}au, A.~Reiserer, N.~Kalb, M.~S. Blok,
  J.~Ruitenberg, R.~F.~L. Vermeulen, R.~N. Schouten, C.~Abell{\'{a}}n, W.~Amaya,
  V.~Pruneri, M.~W. Mitchell, M.~Markham, D.~J. Twitchen, D.~Elkouss, S.~Wehner, T.~H.
  Taminiau, and R.~Hanson.
\newblock {Loophole-free Bell inequality violation using electron spins separated by 1.3 kilometres}.
\newblock {\em Nature}, 526(7575):682--686, oct 2015.

\bibitem{Shalm2015}
L.~K. Shalm et al., 
\newblock {Strong Loophole-Free Test of Local Realism}.
\newblock {\em Physical Review Letters}, 115(25):250402, dec 2015.

\bibitem{Moreau2019}
Paul-Antoine Moreau, Ermes Toninelli, Thomas Gregory, Reuben~S. Aspden,
  Peter~A. Morris, and Miles~J. Padgett.
\newblock {Imaging Bell-type nonlocal behavior}.
\newblock {\em Science Advances}, 5(7), jul 2019.

\bibitem{Ghosh1986}
R.~Ghosh, C.~K. Hong, Z.~Y. Ou, and L.~Mandel.
\newblock {Interference of two photons in parametric down conversion}.
\newblock {\em Physical Review A}, 34(5):3962--3968, nov 1986.

\bibitem{Groblacher2007}
Simon Gr{\"{o}}blacher, Tomasz Paterek, Rainer Kaltenbaek, {\v{C}}aslav
  Brukner, Marek {\.{Z}}ukowski, Markus Aspelmeyer, and Anton Zeilinger.
\newblock {An experimental test of non-local realism}.
\newblock {\em Nature}, 446(7138):871--875, apr 2007.

\bibitem{Everett1957}
Hugh Everett.
\newblock {\,`Relative State' Formulation of Quantum Mechanics}.
\newblock {\em Reviews of Modern Physics}, 29(3):454--462, jul 1957.

\bibitem{Bohm1952}
David Bohm.
\newblock {A Suggested Interpretation of the Quantum Theory in Terms of ``Hidden'' Variables. I}.
\newblock {\em Physical Review}, 85:166, 1952.

\bibitem{Bell2004}
John~S. Bell.
\newblock {\em {Speakable and Unspeakable in Quantum Mechanics}}.
\newblock Cambridge University Press, 1987/2004.

\bibitem{Schroedinger1926a}
E.~Schr{\"{o}}dinger.
\newblock {Quantisierung als Eigenwertproblem}.
\newblock {\em Annalen der Physik}, 79:361--376, 1926.

\bibitem{Schnabel2025epr}
Roman Schnabel.
\newblock {Discovery of entanglement generation by elastic collision to realise
  the original Einstein-Podolsky-Rosen thought experiment}.
\newblock {\em npj Quantum Information}, 11(1):76, may 2025.

\bibitem{Pironio2010}
S.~Pironio, A.~Ac{\'{i}}n, S.~Massar, A.~Boyer de~la Giroday, D.~N. Matsukevich,
  P.~Maunz, S.~Olmschenk, D.~Hayes, L.~Luo, T.~A. Manning, and C.~Monroe.
\newblock {Random numbers certified by Bell's theorem.}
\newblock {\em Nature}, 464(7291):1021--4, apr 2010.

\bibitem{Acin2016}
Antonio Ac{\'{i}}n and Lluis Masanes.
\newblock {Certified randomness in quantum physics}.
\newblock {\em Nature}, 540(7632):213--219, dec 2016.

\bibitem{Heisenberg1927}
W.~Heisenberg.
\newblock {{\"{U}}ber den anschaulichen Inhalt der quantentheoretischen
  Kinematik und Mechanik}.
\newblock {\em Zeitschrift f{\"{u}}r Physik}, 43(3-4):172--198, mar 1927.

\bibitem{Kennard1927}
E.~H. Kennard.
\newblock {Zur Quantenmechanik einfacher Bewegungstypen}.
\newblock {\em Zeitschrift f{\"{u}}r Physik}, 44(4-5):326--352, apr 1927.

\bibitem{Weyl1927}
H.~Weyl.
\newblock {Quantenmechanik und Gruppentheorie}.
\newblock {\em Zeitschrift f{\"{u}}r Physik}, 46(1-2):1--46, nov 1927.

\bibitem{Robertson1929}
H.~P. Robertson.
\newblock {The Uncertainty Principle}.
\newblock {\em Physical Review}, 34(1):163--164, jul 1929.

\bibitem{Planck1900}
Max Planck.
\newblock {{\"{U}}ber das Gesetz der Energieverteilung im Normalspektrum}.
\newblock {\em Annalen der Physik}, 4(4):553--563, 1900.

\bibitem{Heisenberg1925}
W.~Heisenberg.
\newblock {{\"{U}}ber quantentheoretische Umdeutung kinematischer und
  mechanischer Beziehungen}.
\newblock {\em Zeitschrift f{\"{u}}r Physik}, 33:879, 1925.

\bibitem{Born1925}
M.~Born and P.~Jordan.
\newblock {Zur Quantenmechanik}.
\newblock {\em Zeitschrift f{\"u}r Physik}, 34(1):858--888, dec 1925.

\bibitem{Born1926}
M.~Born, W.~Heisenberg, and P.~Jordan.
\newblock {Zur Quantenmechanik. II.}
\newblock {\em Zeitschrift f{\"{u}}r Physik}, 35(8-9):557--615, 1926.

\bibitem{Schroedinger1926}
E.~Schr{\"{o}}dinger.
\newblock {Quantisierung als Eigenwertproblem}.
\newblock {\em Annalen der Physik}, 79:361--376, 1926.

\bibitem{Bohr1928e}
Niels Bohr.
\newblock {The Quantum Postulate and the Recent Development of Atomic Theory}.
\newblock {\em Nature}, 121(3050):580--590, apr 1928.

\bibitem{Ou1992}
Z.~Y. Ou, S.~F. Pereira, H.~J. Kimble, and K.~C. Peng.
\newblock {Realization of the Einstein-Podolsky-Rosen paradox for continuous
  variables}.
\newblock {\em Physical Review Letters}, 68(25):3663--3666, jun 1992.

\bibitem{Schnabel2017}
Roman Schnabel.
\newblock {Squeezed states of light and their applications in laser
  interferometers}.
\newblock {\em Physics Reports}, 684:1--51, apr 2017.

\bibitem{Furusawa1998}
A.~Furusawa, J.~L. S{\o}rensen, S.~L. Braunstein, C.~A. Fuchs, H.~J. Kimble, and E.~S. Polzik.
\newblock {Unconditional quantum teleportation}.
\newblock {\em Science}, 282(5389):706--9, oct 1998.

\bibitem{Bowen2003}
Warwick~P. Bowen, Nicolas Treps, Ben~C. Buchler, Roman Schnabel, Timothy~C.
  Ralph, Hans-A. Bachor, Thomas Symul, and Ping~Koy Lam.
\newblock {Experimental investigation of continuous-variable quantum  teleportation}.
\newblock {\em Physical Review A}, 67(3):032302, mar 2003.

\bibitem{Bowen2003b}
W.P. Bowen, N.~Treps, B.C. Buchler, R.~Schnabel, T.C. Ralph, and T.~Symul.
\newblock {Unity gain and nonunity gain quantum teleportation}.
\newblock {\em IEEE Journal of Selected Topics in Quantum Electronics},
  9(6):1519--1532, nov 2003.

\bibitem{Eberle2013}
Tobias Eberle, Vitus H{\"{a}}ndchen, and Roman Schnabel.
\newblock {Stable control of 10 dB two-mode squeezed vacuum states of light}.
\newblock {\em Optics Express}, 21(9):11546--11553, may 2013.

\bibitem{Julsgaard2001}
Brian Julsgaard, Alexander Kozhekin, and Eugene~S Polzik.
\newblock {Experimental long-lived entanglement of two macroscopic objects}.
\newblock {\em Nature}, 413(6854):400--3, sep 2001.

\bibitem{Peise2015}
J.~Peise, I.~Kruse, K.~Lange, B.~L{\"{u}}cke, L.~Pezz{\`{e}}, J.~Arlt,
  W.~Ertmer, K.~Hammerer, L.~Santos, A.~Smerzi, and C.~Klempt.
\newblock {Satisfying the Einstein-Podolsky-Rosen criterion with massive particles}.
\newblock {\em Nature Communications}, 6(1):8984, dec 2015.

\end{thebibliography}

\end{document}